\documentstyle[preprint,aps]{revtex}

\begin{document}

\title{Relativity in rotating frames}

\author{Asher Peres}
\address{Department of Physics, Technion---Israel Institute of
Technology, 32000 Haifa, Israel}

\maketitle
\begin{abstract}
A common error, originally due to Ehrenfest, is corrected.
\end{abstract}

\bigskip

A recent book \cite{book} with the same title as above contains 17
articles by various authors, following the English translation of a
brief note by Paul Ehrenfest \cite{ehrenfest} who applied a Lorentz
contraction to the rim of a rotating disk, whose radius was not
contracted. This is plainly wrong: Lorentz contractions occur when we
compare the lengths of segments which are at rest in different inertial
frames in relative motion with respect to each other. However, no
two points on the rotating disk have the same velocity \cite{peres}
and therefore they cannot define any Lorentz frame, not even an
``instantaneous'' one.

In the real world, if we want to impart a constant velocity to a
real rod, we can kick it at one end so that the rod acquires a linear
momentum. However there are no rigid bodies, there are only elastic
or plastic ones. The kick will slowly propagate throughout our rod
as elastic vibrations which will gradually damp out, so that the rod
finally imitates a Lorentz frame. Likewise, if we kick a disk to
give it angular momentum, it will gradually approach the limit of
uniform rotation, but then no two points on the disk will have the
same velocity, and the rule for Lorentz contraction is not applicable.

What actually happens is easily seen by using rotating coordinates,
$\phi'=\phi+\omega t$,  in which the disk is ``at rest.''  This is
a passive transformation, a mere relabelling of coordinates, not an
active transformation, and there is no conceptual difficulty. In that
noninertial frame, a centrifugal potential appears, which causes an
elatic distortion of the disk. This is a standard problem of elastic
equilibrium (or of strength of materials). The theory of relativity
is not involved at all.

This work was supported by the Gerard Swope Fund.

\end{document}